\documentclass[aps,prd,nofootinbib]{revtex4-1}
\usepackage{epsfig}
\usepackage{bm}\usepackage{footnote}
\usepackage{graphicx}
\usepackage{amssymb}
\usepackage{amsmath}\usepackage{slashed}
\usepackage{color}
\usepackage{slashed}
\def\p{\psi}

\def\poinc{Poincar\'{e} }
\newcommand{\la}{\langle}
\newcommand{\ra}{\rangle}
\newcommand{\bfr}{{\bf r}}\def\bfQ{\bf Q}

\newcommand{\ben}{\begin{displaymath}}
\newcommand{\een}{\end{displaymath}}
\newcommand{\be}{\begin{equation}}
\newcommand{\ee}{\end{equation}}
\newcommand{\bea}{\begin{eqnarray}}
\newcommand{\eea}{\end{eqnarray}}

\newcommand{\eq}[1]{Eq.~(\ref{#1})}

\newcommand{\bfp}{{\bf p}}

\newcommand{\bfq}{{\bf q}}
\newcommand{\bfk}{{\bf k}} 
\newcommand{\bfb}{{\bf b}}\newcommand{\bfR}{{\bf R}}

\def\g{\gamma}
\def\e{\epsilon}

\def\a{\alpha}
\def\D{\Delta}\def\d{\delta}
\def\r{\rho}\def\t{\tau}\def\P {\Psi}
\def\m{\mu}
\def\n{\nu}
\def\e{\epsilon}
\def\l{\lambda}
\def\G{\Gamma}

\def\s{\sigma}

\def\O{\Omega}\def\vP{\vec P}\def\vq{\vec q}\def\vp{\vec p}
\def\k{\kappa}

\begin{document} 
 
\title{\hskip11cm NT@UW-18-20\\
Defining the Proton Radius: a Unified Treatment } 


\author{Gerald~A.~Miller}
 \affiliation{Department of Physics,
University of Washington, Seattle, WA 98195-1560}
\date{\today}

\begin{abstract}
{\bf Background:} There is significant current interest in knowing the value of the proton radius and also its proper definition. 
{\bf Purpose:} Combine the disparate  literatures of hydrogen spectroscopy and diverse modern parton distributions  to  show that the quantity $r_p^2\equiv -6 G_E'(0)$ is the   relativistically proper
definition that originates from the separate  bodies of work. 
{\bf Methods:} Use   perturbation theory, light-front dynamics and elementary techniques  to find relativistically correct definitions of the proton radius and charge density. 
{\bf Results:}  It is found that the very same proton radius is accessed by measurements of  hydrogen spectroscopy and  elastic lepton scattering.  The derivation of  the mean-square radius as a moment of a spherically symmetric  three-dimensional density is shown to be incorrect.  A relativistically-correct, two-dimensional charge density is related to the diverse modern literature of various parton distributions. Relativistically invariant moments thereof  are derived in a new relativistic moment expansion, the RME.
\end{abstract}
 
 \maketitle 
\noindent

\section{Introduction}
 
What is the value of the radius of the proton? This question has generated much interest since 
 the publication of the results of the  muon-hydrogen spectroscopy  experiment in 2010~\cite{pohl:2010zza} and its confirmation~\cite{Antognini:1900ns}. The proton radius  was measured to be 
$ r_p =0.84184 (67)$   fm, which contrasted with the value obtained from electron-hydrogen spectroscopy  $  r_p =0.8768 (69)$ fm. At that time the large  value was consistent with that obtained (with much larger uncertainties \cite{Horbatsch:2015qda}) from electron scattering.   This   difference of about 4\% has become known as the proton radius puzzle~\cite{Miller:2011yw}.  
The proton radius puzzle was  reviewed  in 2013~\cite{Pohl:2013yb} and 2015~\cite{Carlson:2015jba}. New experimental results for hydrogen  have appeared since  that time   \cite{beyer:2017,Fleurbaey:2018fih} without resolving the puzzle.  More results are planned. The PRAD experiment~\cite{Gasparian:2014rna} seeks to make previous electron scattering determinations of $r_p$ more precise by making measurements at very small values of momentum transfer.  A new measurement of the $2S_{1/2}-2P_{1/2}$ transition in hydrogen is expected to appear (E. A. Hessels, 2016 talk at ECT*). All possible explanations of the proton radius puzzle will be addressed in the Muon-Scattering Experiment  (MUSE) by measuring  $e^\pm-p$ and $\m^\pm-p$ scattering~\cite{Gilman:2013eiv} . \\

One might wonder whether or not a 4\% difference really matters. After all, 4\% is pretty small and (at the present time)  the value of $r_p$ cannot be calculated to that accuracy. Perhaps the most interesting  issue is whether or not the fundamental electron-proton interaction is the same as the muon-proton interaction. To find that this is not the case is to discover a violation of  the  principle of lepton-universality, a cornerstone  of the standard model. \\

But there is another  basic question that must be addressed: what {\bf is} the radius of the proton? How does one define the radius of a quantum-field theoretic system made of nearly massless quarks and gluons? This quantity can be measured in the hydrogen atom and also in electron-proton  scattering. There is a separate, but clear, literature in the fields of atomic and nuclear physics. Both fields obtain the same answer that 
\bea -6 G_E'(0)\equiv r_p^2 ,\label{basic}\eea
where $G_E(Q^2)$ is the Sachs electric form factor. This form factor is a specifically defined   probability amplitude
that an interaction between  a photon of four-momentum $q^\m\,  (Q^2=-q^2)$ and a charged constituent of the proton  can  absorb such a  momentum   with the proton  remaining in its  ground
state. The meaning of \eq{basic} is that the quantity $-6G_E'(0)$ appears in both hydrogen spectroscopy and lepton-proton elastic scattering measurements. The  quantity $r_p^2$ is merely an abbreviation.  The expression \eq{basic} is uniquely used in spectroscopy and scattering experiments to determine the proton radius from the slope of $G_E$.  \\


The aims of the present paper are: 
\begin{itemize}
\item unite  the hydrogen spectroscopy literature  with that of lepton-proton scattering and show how \eq{basic} emerges from the separate bodies of work
\item show that a three-dimensional charge density cannot be defined, so that $r_p^2$ is {\bf not}  a second moment of a density distribution.
\item remind readers how it is that a two-dimensional charge density  which is a matrix element of a density operator between identical initial and final states,
can be defined and determined by the Dirac form factor,  $F_1$
\item place the two-dimensional charge density in the modern context of generalized parton distributions and Wigner functions
\item derive a relativistically correct moment expansion of $F_1$.
\end{itemize}

The second item may be considered controversial by some. This is because the   text-book  interpretation~\cite{BMot,PB,Frauenfelder:1979ii,Cheng:1979ay,Halzen:1984mc,Wong:1998ex,Thomas:2001kw,Close:2007zzd}
 of $G_E$  is  that its 
 Fourier transform is a three-dimensional charge density. 
This interpretation is deeply embedded in the 
 thinking of nuclear and particle physicists 
and therefore continues to guide intuition, as it has since the days
 of the Nobel prize-winning work of Hofstadter~\cite{Hofstadter:1956qs,Hofstadter:1958,RHN}.
 Nevertheless, the relativistic motion of the nearly massless fermionic constituents of the
proton  causes the text-book interpretation 
to be incorrect  because  relativistic invariance is ignored in defining the three-dimensional density.  \\ 

The  modern day literature regarding the measurable aspects of the proton, which is consistent with relativity,  is much deeper than the understanding from 1956. The increasing availability of high energies and high luminosities at fixed target and collider experiments~\cite{Boer:2011fh,Dudek:2012vr}
 allows for unprecedented access to the internal transverse spatial and momentum distributions of   charge distributions inside nucleons and in nuclei.
 The standard framework~\cite{Ji:2003ak} is that of Wigner distributions~\cite{Wigner:1932eb} that allow simultaneous knowledge of both  spatial and momentum aspects of the nucleon wave function. Knowledge of the Wigner distributions allows the construction of 
   generalized parton distributions (GPDs)~\cite{Mueller:1998fv,Ji:1996nm,Ji:1996ek,Radyushkin:1996nd,Radyushkin:1997ki,Collins:1996fb,Ji:1998pc,Radyushkin:2000uy,Goeke:2001tz,Diehl:2003ny,Ji:2004gf,Belitsky:2005qn,Burkardt:2002hr} and  transverse momentum distributions (TMDs)~\cite{Ralston:1979ys,Collins:1981uk,Mulders:1995dh,Boer:1997nt,Belitsky:2002sm,Miller:2008sq} that are generalizations of the usual collinear parton distributions. 
   The variables of the widely used relativistic formalism involve three dimensions-- one is the longitudinal momentum  of a parton and the other two involve either the transverse positions (GPD)  or momenta (TMD). The longitudinal and transverse degrees of freedom are treated separately. This is necessary to maintain symmetries and sum rules provided by relativistic invariance. Electromagnetic form factors may be obtained by doing integrals over  the longitudinal momentum coordinate of GPDs. These form factors must  be described using the same variables as the other observables. Thus, only a two-dimensional charge density may be defined.\\

Let's outline the remainder of this paper.  The appearance  of the proton radius, $r_p$,  in hydrogen spectroscopy  is discussed in Sect.II. It explains how  the key points related to extracting the value of the proton radius were already clearly explained in Refs~\cite{Eides:2000xc,Eides:2007xc,Eides:2014swa}. It is nevertheless worthwhile to repeat,  publicize  this earlier discussion, try to re-emphasize the key points  and strengthen the connection with treatments of  lepton-proton scattering. 
Sect.~III shows that the only existing   derivation of a three-dimensional, spherically symmetric  charge density is faulty.  A properly defined relativistic three-dimensional charge density with  modern formulations is discussed in Sect.~IV. This quantity is intimately connected with  modern formulations of the diverse set of possible parton distributions.
  The ensuing phenomenology is discussed in Sect.~V in which  a  correctly defined moment expansion $\rm RME$ is derived.  Some details are placed in Appendices.

\section{Hydrogen atom}

This section is concerned with understanding the role of the proton radius in hydrogen  spectroscopy. The starting point is to understand
 the leading relativistic corrections to the basic Dirac energy levels. The standard procedure is well-documented in Refs.~\cite{Eides:2000xc,Eides:2007xc}, and their discussion is used here.
In the center of mass system the non-relativistic Hamiltonian for a system of a proton and a lepton (of  mass $m$)  with a  Coulomb interaction is given by
\bea H_0={\vp^2\over 2m}+{\vp^2\over 2M}-{\a\over r}
.\eea
In a non-relativistic loosely bound system an expansion in powers of $\a^2$ corresponds to an expansion in powers of $v^2/c^2$. To proceed one needs an effective Hamiltonian including terms of order $v^2/c^2$. Breit \cite{Breit:1929zz,Breit:1930zza} considered such a Hamiltonian, realizing that all corrections to the non-relativistic Hamiltonian of order $v^2/c^2$ may be obtained  from the sum of the free relativistic Hamiltonians of each of the particles along with relativistic one-photon exchange between the fermions. An explicit expression  for the resulting Breit potential was derived \cite{Barker:1955zz} from the one-photon exchange amplitude using the Foldy-Wouthhuysen transformation. If hyperfine effects are ignored, the result to order $v^2/c^2$  is given by
\bea
V_{\rm Breit}={\pi\a\over 2}\left({1\over m^2}+{1\over M^2}\right)\delta(\vec r)- {\a\over 2mM r}\left(\vp^2+{\vec r (\vec r\cdot\vp)\cdot\vp \over r^2}\right)+{\a\over r^3}\left({1\over 4m^2}+{1\over 2mM}\right)[\vec r\times \vp]\cdot\vec\sigma.
\label{Breit}\eea

All contributions to the energy levels up to order $\a^4$ may be calculated from the total Hamiltonian $H_0+V_{\rm Breit}$. The corrections of order $\a^4$ are the first-order matrix elements of the Breit interaction between the Coulomb-Schroedinger eigenfunctions of $H_0$. The result is 
\bea E_{nj}= m+M -{m_r \a^2\over 2n^2}-{m_r \a^4\over 2n^3}\left({1\over j+{1\over2}}-{3\over4n}+ {m_r\over 4n(m+M)}\right)+{\a^4m_r^3\over 2n^3M^2}\left({1\over j+{1\over2}} -{1\over l+{1\over2}}\right)(1-\delta_{l0}),
\label{energy}\eea
where  the reduced mass $m_r= mM/(m+M)$ and $j$ in the total angular momentum quantum number of the lepton.
Note the presence of the last term of \eq{energy} which removes the degeneracy of the Dirac spectrum between levels with the same $j$ and $l=j\pm {1\over2}$.\\

The expressions \eq{Breit} and \eq{energy} are obtained assuming the proton is a point-like proton. Electromagnetic form factors are introduced to include the effects of  its non-zero spatial extent.   The photon-proton vertex operator  $\G^\m$ is given by
\bea \G^\m=\g^\m F_1(Q^2)+i{\s^{\m\n}\over 2M}\kappa F_2(Q^2),\label{vert}\eea
where $Q^2>0$ is the negative of the square of the virtual space-like photon momentum, $M$ is the proton mass, $F_1$ is the Dirac form factor, $F_2$ is the Pauli form factor and $\k$ is the proton anomalous magnetic moment. It is useful to define the Sachs form factors:
\bea 
G_E(Q^2)=F_1(Q^2)-\t \k F_2,\,G_M(Q^2)=F_1(Q^2)+\k F_2(Q^2),\label{GEdef}\eea
 where $\t\equiv {Q^2\over 4M^2}$. With this notation $F_{1,2}(0)=1$.  \\

The photon-electron vertex function in a hydrogen-like atom is given~\cite{Eides:2000xc,Eides:2007xc} as the matrix element: $\bar u(\vp',s')\G^\mu u(\vp,s) $ with spinors normalized as $u^\dagger u=1$.
Calculation, see {\it eg.}~\cite{Eides:2000xc,Eides:2007xc,Bertozzi:1972jff}, to order $1/M^2$  (and ignoring a spin-orbit term) reveals that
\bea \bar u(\vp',s')\G^0u(\vp,s)= (1-{\vq^2\over 8M^2})G_E(\vq^2) 
\label{key}\eea
This  key equation is derived in Appendix A. The left-hand side of \eq{key} is the time component of a four-vector. The right-hand-side does not depend on $\vp$; it is frame-independent, and may be used to identify the leading proton-radius effect.

One  uses the lowest order Taylor expansion, keeping only the spin-independent term, to extract the proton radius from the measured energy levels. Thus one writes:
\bea G_E(\vq^2)=1 +\vq^2 G_E'(0).\label{slope}
\eea
{The difference between $Q^2$ and $\vq^2$, $-q_0^2,$ is of order $\a^2m^2/M^2$ and is a higher order correction, and 
any  correction arising from a $G_E''(0)$ term is completely negligible~\cite{Stryker:2015ika}.
We then find (to order $\vq^2$) that
\bea \bar u(\vp',s')\G^0u(\vp,s)\rightarrow  
1-{\vq^2\over 8M^2}+ \vq^2 G_E'(0).\eea
    The  term  ${\vq^2\over 8M^2}$ leads to the   Darwin term in the lepton-proton interaction   \cite{Barker:1955zz} that provides  the term proportional to $\d_{l0}$ in \eq{energy}, so it is already included. The net result is that the effect of the proton  size is given simply  by
including the term $ \vq^2G_E'(0)$ in the lepton-proton vertex function.\\

Keeping the non-zero size of the proton leads to  the  momentum space  version of the Coulomb potential: 
\bea V_C(\vq^2)=-{4\pi\a\over \vq^2}(1+\vq^2G_E'(0))= -{4\pi\a}({1\over  \vq^2}+G_E'(0))\eea
The coordinate space potential $V_C(r)$ is given by  the three-dimensional Fourier transform:
\bea V_C(r)= -\int {d^3q\over (2\pi)^3} e^{-i\vec q\cdot \vec r}{4\pi\a}({1\over  \vq^2}+G_E'(0))=-{\a\over r}-{4\pi\a}G_E'(0)\d(\vec r).
\eea 
Since  $G_E$ falls with increasing $\vq^2$
 one finds a repulsive correction to the Coulomb potential, $\D V_C$, given by
\bea \D V_C(\vec r)=-{4\pi\a}{ G_E'(0)}\d(\vec r).\label{dv}
\eea\\

The typical value of $\vq^2$ is of the order of the square of the inverse of the Bohr radius of the atom. The muonic hydrogen atom Bohr radius is about 200 times smaller than  that for the electronic one. This huge difference does not influence the potential 
 $\D V_C$ because of the cancellation of  the factor $\vq^2$ by its inverse that arises from the photon propagator. The net result is the delta function appearing in \eq{dv}.  The difference between Bohr radii would enter if one included the $\vq^4$ term in the Taylor expansion of $G_E$, but such terms are smaller by the ratio of the proton size to the Bohr radius~\cite{Stryker:2015ika}.\\

The shift in the energy, $\D E$,  is given by the matrix element:
\bea \D E=\la \p_{nl} |\D V_C|\p_{nl}\ra=
 -{4\pi}\a G_E'(0) |\p_{n0}(0)|^2\d_{l0}.\label{good}\eea
The net result is that the energy shift in the hydrogen atom is determined by the slope of $G_E$  at its origin.  The effect, \eq{good}, is of order $\a^4$ because $ |\p_{n0}(0)|^2$ is of order $\a^3$, so that this term should be included with the others of order $\a^4$  displayed in \eq{energy}.  Effects of higher order in $\a$ are not considered here.\\

For historical reasons, to be discussed in the next Section, the slope is redefined as 
\bea G'_E(0)=-{r_p^2\over 6},\eea   with 
\bea G_E(Q^2)=1-{r_p^2\over6}\vq^2\eea for sufficiently small values of $\vq^2$. This means that 
one may also write
\bea \D E={4\pi}\a {r_p^2\over 6} |\p_{n0}(0)|^2\d_{l0},\eea
as is often done.

\section{Lepton-proton scattering}

The electron-proton elastic scattering cross section, obtained under the assumption that the lepton-proton interaction is mediated by a single photon (and neglecting the electron mass) is most simply expressed in terms of $G_E,G_M$~\cite{Rosenbluth:1950yq}:
 \bea
 {d\s\over d\O}=\left({d\s\over d\O}\right)_M \times[G_E^2+{\tau\over\e}G_M^2]{1\over 1+\tau},\eea
 where $\left({d\s\over d\O}\right)_M$ is the Mott cross-section in which the proton is treated as point-like ($F_{1,2}(Q^2)=1)$, and $\e$ is a kinematic factor.
 Including the non-zero value of the muon mass leads to a slightly more complicated expression~\cite{Preedom:1987mx}.
 \\
 
Elastic electron-proton scattering experiments were first performed at  Stanford by Hofstadter and collaborators, and summarized in
Ref.~\cite{Hofstadter:1958}. This early work, which famously~\cite{RHN} discovered that the proton was not a point-particle,   assumed that $F_1=F_2\equiv F$. Their analysis used the equations:
\bea
\lim_{Q_L^2\to0} F(Q_L^2)=1-{Q_L^2a^2\over 6}+\cdots,\eea in which $a$ was associated with the physical extent of the proton and  $Q_L$ is the {\it  laboratory value} of  the magnitude of the three-momentum transfer. It is further asserted that
\bea 
F(Q_L^2)=\int_0^\infty \r(\bfr)e^{i \vec Q_L\cdot\bfr}d^3r,\label{wrong}\eea
where
The authors are very careful about using this expression. They state that \eq{wrong} {\it ``applies in the non-relativistic limit in which $\r(\bfr)$ is the static  density distribution"}. This definition is frame-dependent and so violates the principle of relativity.
\\

Expansion of the exponential appearing  in \eq{wrong}  leads to the well-known moment expansion:
\bea F(Q_L^2)=1- {Q_L^2\over6}\la r^2\ra +{Q_L^2\over 120} \la r^4\ra+\cdots\label{nrmom}\\
\la r^n\ra =\int r^n\r(\bfr)d^3r.\eea The validity of this expansion depends upon the validity of the non-relativistic limit.  But there is {\bf no} reason to believe that any non-relativistic treatment is valid for treating the proton form factor because elastic  electron-proton scattering proceeds mainly via the absorption of a virtual photon by a nearly massless up or down quark.\\

Sachs {\it et. al.}~\cite{Ernst:1960zza,Sachs:1962zzc}  introduced the so-called  Breit frame in which $q^0=0$ so that here (and in all following equations)  $Q^2$ is the  Lorentz scalar quantity, $Q^2=-q^2$, and $q^\m$ is the four-momentum  of the  single-photon mediator. Sachs~\cite{Sachs:1962zzc}  argued that in this frame the charge  density is given by the Fourier transform of $G_E$. The resulting  non-relativistic (NR) density as $\r_{\rm NR}(r)$ is defined by the equation: 
\bea
\r_{\rm NR}(r)\equiv \int {d^3Q \over (2\pi)^3}e^{-i\bfQ\cdot\bfr}G_E(Q^2).\label{rhoNR}
\eea
 \\

We shall show that the   definition \eq{rhoNR}  (despite its wide use) has no connection with well-defined matrix elements of  quantum field theory.  Equation~(\ref{rhoNR}) can be used to obtain the results that
\bea
\la r^2\ra_{\rm NR}\equiv \int d^3r r^2\r_{\rm NR}(r)
=-6\, G_E'(0)\label{r2nr}\\
\la r^4\ra_{\rm NR}\equiv \int d^3r r^4\r_{\rm NR}(r)=60\, G_E''(0)
.\eea\\

\eq{r2nr}  has often been    used to analyze the charge distribution of the neutron $n$. The result is that the  mean square charge radius, $\la r^2_n\ra_{\rm NR}$,  is almost completely accounted for numerically  by the anomalous magnetic term, $3\kappa_n/(2M^2)$,  arising from the $F_2$  contribution to $G_E$~\cite{Thomas:2001kw}. This seems very  strange and looks like a puzzle. The puzzle dissipears if one realizes that the slope of $G_E$ is not related to the expectation value of $r^2$ in a spherically symmetric charge density. The pion form factor is another example of oddity.  The 3-dimensional Fourier transform of the monopole form factor (that approximates  the data) is singular at the origin \cite{Miller:2009qu}. \\

\begin{figure}[h]
\centering
\includegraphics[scale = 0.34]{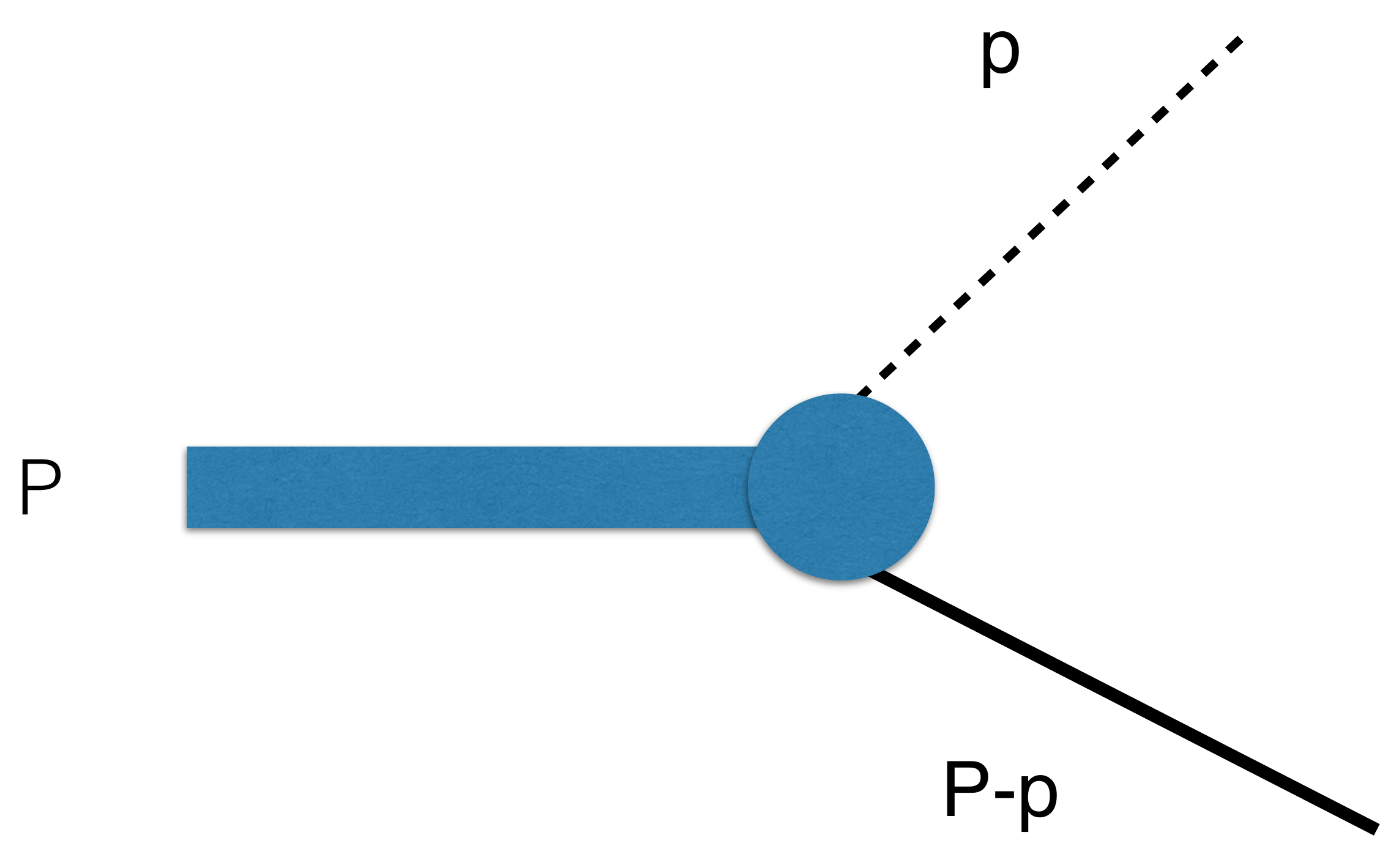}
\caption{[Color online] Bethe-Salpeter wave function. A hadron of momentum $P$ fluctuates into a constituent of four-momentum $p$ and another of momentum $P-p$}
\label{BSWF}\end{figure}

There is a more technical way of explaining why the non-relativistic approach fails.
The difficulty comes because proton wave functions of differing momentum are different. These functions are related by a boost, which in general contains an exponential of an operator as complicated as the strong interaction Hamiltonian~\cite{Gasiorowicz:1966xra,Cahill:2013pez}. To see this,  consider a very  simple example in which a scalar proton of mass $M$ is made of a scalar quark (of mass $m_q$ and a scalar di-quark (of mass $M_S$).   In this case the 
 Bethe-Salpeter  wave function is given by
\bea \p_{\rm B.S.}(P,p) ={1\over p^2-m_q^2+i\e}{1\over (P-p)^2-M_S^2+i\e}={1\over p^2-m_q^2+i\e}{1\over M^2-2 P\cdot p +p^2-M_S^2+i\e},
\eea
in which various constants have not been displayed. This wave function is shown in  Fig.~\ref{BSWF}.\\

Boosting the wave function in this simple  model is achieved merely by changing the value of the four-momentum $P$. 
In lepton-proton elastic scattering the four momentum of the  initial   proton is $P$ and that of the final one  is $P'=P+q$. 
The initial-state  wave function depends upon  $P\cdot p$, which in the laboratory frame is given by $Mp^0$. The final-state wave function depends on $P'\cdot p=(P+q)\cdot p= (M+q^0)p^0-\bfq\cdot\bfp$. Thus electron scattering involves two different wave functions. A density requires the appearance of the square of a wave function, which does not appear here.\\

\subsection{Breit Frame Falibility}
The Breit frame (introduced in Ref.~\cite{Ernst:1960zza})  is the one in which the three-momentum of the initial proton, $\vP$, is -1/2 that of the incident virtual photon, $\vq$,     $\vec P =-\vec q/2$. The final proton momentum $\vP'=\vec P +\vec q=\vec q/2,$  and   the initial and final protons have the same energy. However, the initial and final wave functions are different because the quantities $P\cdot p$ and $P'\cdot p$ must appear in any relativistic wave function and
\bea P\cdot p =\sqrt{M^2+\bfq^2/4}p^0+\bfq\cdot\bfp/2,\, P'\cdot p =\sqrt{M^2+\bfq^2/4}p^0-\bfq\cdot\bfp/2\eea
These differ, so that again  one is not dealing with the square of a given wave function. \\

Ref.~\cite{Ernst:1960zza}   shows that the matrix element of the time component of $\G^\m,\, \Gamma^0$ is proportional to $G_E$ if the helicity is changed (spin direction is not changed).  
This argument is also given in  {\it e.g.} \cite{Halzen:1984mc}.  In that reference a quantity $\r$ is {\it defined} as the matrix element of $\G^0$:
\bea \r\equiv \bar u(\vec q/2,s_z)\G^0u(-\vec q/2, s_z)= G_E(Q^2)
,\label{BF}\eea
with the normalization again being $u^\dagger u=1$, and where the direction of $\vq$ defines the $z$-axis. The equality follows from evaluating the spinor matrix element. However, {\it no spatial dependence is implied by this definition of $\r$}. 
\\

To the best of my knowledge the derivation of a relationship between 
 $G_E$  and a  three-dimensional charge density is attempted only in Appendix II of Ref.~\cite{Sachs:1962zzc}. In that formalism the initial and final states are ``{\it brought to rest}'' through the use of narrow wave packets. 
 This is an attempt to avoid the previously mentioned problem associated with the boosts. \\

Although Ref.~\cite{Sachs:1962zzc} made a strong attempt to derive the charge density as 
 a three-dimensional Fourier transform of $G_E$,  the derivation is simply wrong.  To show this,    I redo the calculation of Appendix II  of that paper, avoiding    an incorrect assumption used there.\\
 
 Sachs writes the proton wave packet  state as
 \bea|\P\ra=\int d^3P g(\vec P)|P,s\ra,\label{S1}\eea 
 with $g(\vec P)$ representing a narrow wave packet.   The function $g(\vec P) $ is sufficiently narrow so that
\bea |g(\vec P)|^2\rightarrow \delta (\vec P).\label{S2}\eea It is worthwhile to use an  explicit representation. Using a Gaussian
\bea g(\vec P)={ R^{3/2}\over \pi^{3/4}} \exp{[-\vP^2 R^2/2]},\label{gdef}\eea
where $R$ is assumed to be infinite, as a first choice is convenient. The  meaning of the expression \eq{gdef}  is the standard one of a distribution  in which  the first step is to do all of the relevant  integrals, keeping $R$ finite. Then,  with the values of the integrals in hand  take  $R$ to infinity~\cite{Born:1999ory,AW2005}.\\

Sachs proceeded by computing moments  of the putative charge distribution. Only  the time-component, and a quadratic moment is relevant  to display  the so-called charge density.  In this case:
\bea M^{(2)}\equiv \int d^3q\int d^3p\, g^*(\vec p+\vec q/2)g(\vp-\vec q/2)\int d^3x \,x^2 \la P',\l|j^0(x)|\vP,\l\ra,\label{mom2}\eea
with the  integration variables   chosen as $\vp $ and $\vec q$ with $\vP'=\vp+\vec q/2, \vP=\vp-\vq/2$, and  $x^2\equiv \sum_ix_i^2$  with $x_i$ as  the Cartesian component of the three vector $\vec x$.  \\

The matrix element of the time-component of the current is given by
\bea
\la P'|j^0(x)|P\ra=(2\pi)^{-3} e^{iq\cdot x} \bar u(\vec P',\l)\G^0u(\vP,\l),\eea
which (after replacing $x^2$ by $-\nabla_q^2$,  integration over $\vec x$, and integration by parts) leads to the expression:
\bea M^{(2)}= - \int d^3q\d(\vec q) { {\nabla_q }^2}\int d^3p\, g^*(\vp+\vec q/2)g(\vp-\vec q/2)\bar u(\vp+\vq/2,\l)\G^0u(\vp-\vq/2,\l)e^{iq^0 t}.\label{M20}
\eea
At this stage Sachs made the replacement $ g^*(\vec p+\vec q/2)g(\vp-\vec q/2)\rightarrow |g(p)|^2$. The justification is that ``{\it terms resulting from the shape of the wave packet are not of interest here and are therefore dropped.}". However, this is not a correct justification for the replacement because (as a $\d$ function) the quantity  $|g(P)|^2 $ must vary  rapidly. One needs to be careful about the derivatives.   To see this, use the specific form of \eq{gdef} in \eq{M20}. Then 
\bea &M^{(2)} = - \lim_{R\to \infty}\int d^3q\d(\vec q) {\nabla_q^2 }\int d^3p {R^3\over \pi^{3/2}}\exp{(-\vp^2R^2 -\vq^2R^2/4})\bar u(\vp+\vq/2,\l)\G^0u(\vp-\vq/2,\l)e^{iq^0t}.\label{M2}\eea
The term ${R^3\over \pi^{3/2}}\exp{(-\vp^2R^2)}$ leads to a delta function setting $\vp=0$, so that $q^0=0$ and the Breit-frame result, \eq{BF},  for  the matrix element of $ \G^0$ may be used. Then:
\bea &
 M^{(2)}_i 
=- \lim_{R\to \infty}\int d^3q\d(\vec q) {\nabla_q^2}\,e^{-\vq^2R^2/4}G_E(\vq^2)\\&=- \lim_{R\to \infty}\int d^3q\d(\vec q) \left(-R^2/2 G_E(\vq^2))+ {\nabla_q^2 }G_E(\vq^2)\right)\\&
= \lim_{R\to \infty}(R^2/2 - {\nabla_q }^2G_E(\vq^2))|_{\vq^2=0} =\infty.\label{M21}
\eea
This means  that the quadratic moment is actually infinite!  This moment expansion fails. The underlying reason is that $\nabla_q^2$ must involve the square of some distance and the  infinite parameter $R^2$ must appear in addition to any length scales in the proton.\\

 If one asserts that $G_E(\vq^2)=\int d^3x e^{-i \vq\cdot\vec x} \r_{\rm NR}|(|\vec x)|$, then one finds 
\bea -{\nabla_q }^2G_E(\vq^2))|_{\vq^2=0} =\int d^3x \,x^2  \r_{\rm NR}(|\vec x|),\eea
which looks like the expressions of the usual literature. However, this term (which does appear in \eq{M21}) is overwhelmed by the infinite term that also appears. Note that the infinite result 
  does not rely on using  the specific Gaussian form of \eq{gdef}. It would occur with any specific representation of a delta function, as shown in Appendix B. Thus the derivation of Sachs is fatally flawed.\footnote{The evaluation of \eq{M2} proceeded by first
  obtaining $\d(\vec p) $ and then handling the dependence of $\vq$. The same result, \eq{M21} is obtained if one first differentiates with respect to $\vq$.}

\smallskip
One may understand the failure of the Sachs procedure in simple terms. The wave function $|\P\ra$ is meant to represent a proton  of 0 three-momentum, so that (via the  uncertainty principle)  its position is totally undetermined. This is the origin of the infinite result. Another procedure would be to use \eq{gdef} in the  opposite limit that $R$ is very, very small. This would lead to a wave packet that is concentrated in a narrow region of space, taken as the origin. However, the use of such a wave packet in \eq{mom2} would not allow the use of the Breit frame result
because the integral over $\vp$ would go over all of its values.\\

The net result of all of this is that the relation between $G_E'(0)$ and $r_p^2$,  \eq{basic},  is merely a definition.

\section{True Charge density}


A proper determination of a 
 charge density requires the measurement of a matrix element of a 
 density operator taken between initial and final states that are the same.
 We show here that  the proton form factor, $F_1$
is  a specific  integral of the  
three-dimensional charge
density of partons 
in the infinite momentum frame,  $\hat{\rho}_\infty(x^-,\bfb)$.\\

It is necessary to provide  a brief introduction to light- front coordinates.
 Instead of the usual $x^0=ct,x^3=z$, the light front approach  uses $x^\pm=(x^-\pm x^3)/\sqrt{2}.$ By convention the term $x^+$ corresponds to the time and $x^-$ corresponds to the longitudinal distance coordinate.\\

In the infinite momentum frame, IMF, the time coordinate $ct=x^0/\sqrt{2}$
is expressed in a frame moving along the negative $z$ direction with a
velocity nearly that of light using the Lorentz transformation as the
 variable $x^+=(x^0+x^3)/\sqrt{2}$, with the usual $\gamma$ factor  
 absorbed by a Jacobean of an integral over volume
 \cite{Susskind:1968zz}.
 The
 $x^+$ variable is canonically conjugate to the minus-component of the
 momentum operator $p^-\equiv(p^0-p^3)/\sqrt{2}$. The longitudinal 
spatial variable  is  $x^-=(x^0-x^3)/\sqrt{2}$ and its conjugate
momentum is  $p^+=(p^0+p^3)/\sqrt{2}$. It is this plus-component of
momentum that is associated with the usual Bjkoren variable. The
transverse coordinates $x,y$ are written as $\bfb$ with the conjugate momentum denoted
$\bfp$.  Boldface is used here to denote the  two-dimensional transverse components of position and
momentum vectors to distinguish these   from the three-dimensional vectors  ({\it e.g. $\vq$}) of previous sections.\\

 Light-front time-quantization, which sets $x^+$ and the plus-component of all spatial variables to zero, is used. This means that  $x^-$ can be thought of as the longitudinal variable $-\sqrt{2}x^3.$
One extremely useful aspect of using these
variables  is that Lorentz transformations to frames moving with different 
{ transverse} velocities do not depend on interactions. These
transformations form the kinematic subgroup of the \poinc group, so
 that boosts in the transverse direction are accomplished 
as in the
non-relativistic theory; the dependence on the total transverse momentum  of any system appears only as an overall phase factor. \\

 This language may seem a bit abstract. All it means the wave function of a proton with a given $(p^+,\bfp)$ is related to the one of momentum ($p^+,{\bf0})$ by a factor that is independent of the relative momenta of the partons that make up the wave function.
The necessary integrations to compute form factors (in a frame in which $Q^2=\bfq^2$)   only involve the  relative variables  that appear in light-front wave functions.
Examples can be found in \cite{Brodsky:2000ii,Miller:2009sg,Miller:2010nz}.\\

The density that is relevant here has been known for a long time      \cite{Soper:1976jc} and often been exploited~\cite{Miller:2010nz, Miller:2009qu,Carlson:2007xd}.
In the IMF, the electromagnetic charge density 
$J^0$ operator becomes $J^+$ and
\bea \hat{\rho}_\infty(x^-,\bfb)=J^+(x^-,\bfb)=\sum_q e_q \overline{q}(x^-,\bfb)\gamma^+q(x^-,\bfb)=\sum_q e_q   \sqrt{2} 
q^\dagger_+(x^-,\bfb)q_+(x^-,\bfb),\label{imfop}\eea
where  $q_+(x^\mu)=
\gamma^0\gamma^+/\sqrt{2} q(x^\mu)$, the independent part of the
 quark-field operator $q(x^\mu)$.
The 
  time variable, $x^+$ is set to  zero. Note the appearance of 
 the absolute square
  of quark field-operators, which is the signature of a true density operator.  An analogous expression is widely used to describe color charge densities~\cite{McLerran:1993ka,Dumitru:2018vpr}.\\

The purpose of this section is to show how matrix elements of $ \hat{\rho}_\infty(x^-,\bfb)$ (which are true densities)  emerge from modern quantum field theory treatments of nucleon structure.  
 The vast literature concerning 
the  diverse set of functions that are used to
describe nucleon structure  includes
generalized parton distributions  GPDs \cite{Mueller:1998fv,Ji:1996nm,
Radyushkin:1997ki,Collins:1996fb,Ji:1998pc,Radyushkin:2000uy,Goeke:2001tz,Diehl:2003ny,Ji:2004gf,Belitsky:2005qn,Hagler:2004yt,Boffi:2007yc}
transverse momentum distributions (TMDs)
\cite{Collins:1981uw,Ralston:1979ys,Belyaev:1988xu,Anselmino:1994gn,Mulders:1995dh,Chibisov:1995ss,Pasquini:2009eb}
and, more recently generalized transverse momentum distributions (GTMDs)~\cite{Lorce:2011kd,Echevarria:2016mrc}.\\

Generalized parton distributions   
are of high current interest because
 they 
can be related to the total angular
momentum carried by quarks in the nucleon and can be determined using
deeply virtual Compton scattering experiments~\cite{Ji:1996nm}.  The opportunity  of determining all of these is greatly enhanced by the possible creation of an electron-ion collider~\cite{Accardi:2012qut}.\\


   These distributions are specific  matrix elements of quark-field operators,
between nucleon states, which in contrast to the usual quark distribution functions,
 do not necessarily have the same momenta. The specific case in which
the longitudinal momentum transfer  vanishes,  and the initial and final states have the same helicity  $\lambda'=\lambda$ is relevant in the present context. Then,  
in the light-cone gauge, $A^+=0$, the matrix element defining the GPD, $H_q$ for a quark of flavor $q$ is
\bea&& 
{ H}_{q}(x,t)
=\int\!\! \frac{dx^-}{4\pi}\langle p^+,\bfp',\lambda|
\bar{q}(-\frac{x^-}{2},{\bf 0})
\gamma^+ q(\frac{x^-}{2},{\bf 0})
|p^+,\bfp,\lambda\rangle e^{ixp^+x^-}.
\label{eq:pd1}
\eea
where  the  normalization is  
$\langle {p'}^+,\bfp',\l| {p}^+,\bfp,\l\rangle
=2p^+(2\pi)^3  \delta({p'}^+-p^+)\delta^{(2)}({\bfp}'-\bfp)$.  The variable $\l$ denotes the helicity, and  only the  helicity non-flip  term needed to compute $F_1$  appear here.   
 The four-momentum transfer $q_\alpha=p'_\alpha-p_\alpha$ is  space-like, with 
the square of the space-like four-momentum transfer $q^2=-Q^2$  
and  use  the Drell-Yan (DY) frame with 
$ (q^+=0,Q^2=\bfq^2)$. No longitudinal momentum is transferred,  so
that initial and final states are related only by kinematic
transformations. Moreover,  the current operator links Fock-state components with the same number of constituents.
The abbreviation  
$ -t=-(p'-p)^2=(\bfp'-\bfp)^2=-q^2=Q^2$ is used.
The presence of the operator $\gamma^+$ insures that independent field operators appear 
in the matrix element.\\

GPDs allow for a unified description of a number of
hadronic properties \cite{Ji:1996ek}. Notice  
that if $t=0$ 
 they reduce to conventional PDFs
$
H_q(x,0)= q(x)$, and,  of most relevance here, that the 
  integration of $H_q$ over $x$ yields the nucleon 
electromagnetic  form factor:
\be
F_1(t)=\sum_q e_q \int dx H_q(x,t),
\label{eq:form}
\ee
with the defining equation 
\bea F_1(Q^2)={\langle {p'}^+,\bfp',\l|J^+(0)|p^+,\bfp,\l\rangle\over 2p^+}.\label{fdef}\eea

The spatial structure of a  nucleon can be examined if one uses the fact that  transverse boosts are independent of interactions in the infinite momentum frame \cite{Kogut:1969xa,Burkardt:2005td} to 
define   \cite{Soper:1976jc,Burkardt:2002hr,Diehl:2002he} nucleonic 
states that are transversely localized.  The state with transverse center of mass
$\bfR$ set to 0 is formed by taking a  linear superposition of
states of transverse momentum.
In particular,
\be
\left|p^+,{\bf R}= {\bf 0},
\lambda\right\rangle
\equiv {\cal N}\int \frac{d^2{\bf p}}{(2\pi)^2 \sqrt{2p^+}} 
\left|p^+,{\bf p}, \lambda \right\rangle,
\label{eq:loc}
\ee
where $\left|p^+,{\bf p}, \lambda \right\rangle$
are light-cone helicity eigenstates
\cite{Soper:1976jc} and
${\cal N}$ is a normalization factor satisfying
$\left|{\cal N}\right|^2\int \frac{d^2{\bf p}_\perp}{(2\pi)^2}=1$.
References~\cite{Burkardt:2000za,Diehl:2000xz} use  
wave packet treatments that  avoid states 
normalized to $\delta$ functions, but  this  leads to the
same results as using \eq{eq:loc}. Note however,  the relevant range of integration
in \eq{eq:loc} must be restricted to $|\bfp|\ll p^+$ to maintain the interpretation
of a nucleon moving with well-defined longitudinal momentum~\cite{Burkardt:2000za}. Thus  the infinite momentum frame,  with $p^+$ as the large momentum, is  used. 
 This is a frame in which 
the interpretation of a nucleon as a  set of a large number of partons is valid.  \\

Using  \eq{eq:loc} sets 
the  transverse 
center of momentum of 
a state of  total very large 
momentum $p^+$  to zero, so that a
transverse distance $\bfb$ relative to $\bfR$
can be  defined. 
To use this feature   generalize the quark-field operator appearing in  \eq{eq:pd1}
 by making a translation:
\be
\hat{O}_q(x,{\bf b}) \equiv
\int \frac{dx^-}{4\pi}{q}_+^\dagger
\left(-\frac{x^-}{2},{\bf b} \right) 
q_+\left(\frac{x^-}{2},{\bf b}\right) 
e^{ixp^+x^-}.
\label{eq:bperp}
\ee The  
impact parameter dependent PDF is defined \cite{Burkardt:2000za} as the matrix element
of this operator in the state of \eq{eq:loc}:
\be
q(x,{\bf b}) \equiv 
\left\langle p^+,{\bf R}= {\bf 0},
\lambda\right|
\hat{O}_q(x,{\bf b})
\left|p^+,{\bf R}= {\bf 0},
\lambda\right\rangle. 
\label{eq:def1}
\ee\\

The use of \eq{eq:loc} in \eq{eq:def1} allows one to show that 
$q(x,{\bf b})$ is the two-dimensional Fourier transform of the GPD $H_q$:
\bea q(x,{\bf b})=\int {d^2q\over (2\pi)^2}e^{-i\bfq\cdot\bfb}H_q(x,t=-\bfq^2),\label{ft1}
\eea with $H_q$ appearing because the initial and final helicities are each $\lambda$.
A complete determination of $H_q(x,t)$ (with $t\le0$) would  determine
$q(x,{\bf b})$.\\

One finds a probability interpretation \cite{Soper:1976jc} by integrating $q(x,{\bf b})$
over all values of $x$. This  sets the differences in longitudinal distances, appearing in \eq{eq:bperp}, to 0. Then the use of translational invariance leads to the result
\bea
\int dx\;q(x,{\bf b})
=\left\langle p^+,{\bf R}= {\bf 0},
\lambda\right|q_+^\dagger(x^-,\bfb)
q_+(x^-,\bfb)
\left|p^+,{\bf R}= {\bf 0},
\lambda\right\rangle.
\label{den01}
\eea
 This equation shows that the matrix element of a true density operator (square of a quark-field operator) taken between identical initial and final states is experimentally accessible.\\
 
Furthermore, multiplying \eq{den01}
   by  the quark charge $e_q$ (in units of $e$),
sums over quark flavors,  uses \eq{eq:loc} with
$\hat{O}_q(x,{\bf b})=e^{-i\hat{\bfp}\cdot\bfb}\hat{O}_q(x,{\bf 0}) e^{i\hat{\bfp}\cdot\bfb}$ 
 along with \eq{eq:form}, the resulting infinite-momentum-frame  IMF parton
charge density in transverse space
is 
\bea
\rho(b)\equiv \sum_q e_q\int dx\;q(x,{\bf b})=\int {d^2q\over (2\pi)^2} F_1(Q^2=\bfq^2)e^{-i\;\bfq\cdot\bfb}.
\label{rhob0}\eea\\

This relation shows that a properly-defined charge density, the transverse charge density,  is obtained using the same formulations  that is used to define generalized densities. \\

\newcommand{\uD}{\mathcal{D}}
 \newcommand{\ud}{\mathrm{d}}
 \subsection{Wigner distributions}
 There is now a broader perspective  involving  a diverse set of distributions that can be used to characterize  nucleon structure~\cite{Lorce:2011kd}. This subsection is intended to place the transverse density in the context of Wigner distributions.
Wigner distributions in QCD were first explored in Refs.~\cite{Ji:2003ak,Belitsky:2003nz}. Neglecting relativistic effects, those authors used the standard three-dimensional Fourier transform in the Breit frame and introduced six-dimensional Wigner distributions (three position and three momentum coordinates). The modern perspective involves  instead five-dimensional Wigner distributions (two position and three momentum coordinates) as seen from the infinite momentum frame (IMF). These three momentum variables of a quark are $k^+,\bfk$, so there is  no spherically-symmetric charge density.   These light front variables were  exploited~\cite{Lorce:2011kd}  to arrive at a definition of Wigner distributions that is completely consistent with relativity.\\

The first step  is to  use Wigner operators for quarks of flavor $q$ at a fixed light-cone time $y^+=0$:
\begin{equation}\label{wigner-operator}
\widehat W_q(\bfb,\bfk,x)\equiv\frac{1}{2}\int\frac{\ud z^-\,\ud^2z}{(2\pi)^3}\,e^{i(xp^+z^--\bfk\cdot{\bf  z})}\,\overline{q}(y-\tfrac{z}{2})\g^+\mathcal W\,q(y+\tfrac{z}{2})\big|_{z^+=0},
\end{equation}
with $y^\mu=[0,0,\bfb ]$, $p^+$ is the  average of the initial and final nucleon longitudinal momentum and $x=k^+/p^+$ is  the average fraction of nucleon longitudinal momentum carried by the struck quark. The above equation is a specific Wigner operator that involves $\g^+$ that is relevant here. More generally one could use  any twist-two Dirac operator $\Gamma=\gamma^+,\gamma^+\gamma_5,i\sigma^{j+}\gamma_5$ with $j=1,2$. A Wilson line, $\mathcal W$, ensures the color gauge invariance of the Wigner operator by  connecting the points $(y-\tfrac{z}{2})$ and $(y+\tfrac{z}{2})$ see {\it e.g.}  \cite{Meissner:2009ww}. \\\

  Wigner distributions are defined as  matrix elements of the Wigner operators sandwiched between nucleon states with polarization $\vec S$: 
\begin{equation}\label{wigner}
\rho_q(\bfb,{\bf k},x,\l)\equiv\int\frac{\ud^2\Delta}{(2\pi)^2}\,\langle p^+,\tfrac{\bf \Delta}{2},\vec S|\widehat W_q(\bfb,{\bf k},x)|p^+,-\tfrac{\bf\Delta}{2},\vec S\rangle.
\end{equation}
\\

Ref.~\cite{Lorce:2011kd} shows how four different three-dimensional densities can be defined. The task here is to connect  the transverse density, $\r(b)$, of  \eq{rhob0} with the Wigner distribution defined above.  This is done by integrating the quantity $\rho_q(\bfb,{\bf k},x,\l)$  over all values of $x$ and $\bfk$. This sets $z^-$ and $\bf z$ to 0, so that the Wilson line becomes unity, and the result is
\bea  \int dx\, d^2k\langle p^+,\tfrac{\bf \Delta}{2},\l | \widehat W_q(\bfb,\bfk,x)|p^+,-\tfrac{\bf\Delta}{2},\l\rangle=\frac{1}{2p^+} \langle p^+,\tfrac{\bf \Delta}{2},\l |\overline{q}(x^-=0,\bfb)\,\g^+\,q(x^-=0,\bfb)|p^+,-\tfrac{\bf\Delta}{2},\l\rangle,
\eea
where the polarization vector $\vec S$ has been  set to the light front helicity $\l$. Using  translational invariance in the transverse direction and \eq{wigner} shows that
\bea  \int dx\, d^2k\rho_q(\bfb,{\bf k},x,\l )=\int {d^2\Delta\over(2\pi)^2}\frac{e^{-i\bf\Delta \cdot\bfb }}{2p^+} \langle p^+,\tfrac{\bf \Delta}{2},\l |\overline{q}(0)\,\g^+\,q(0)|p^+,-\tfrac{\bf\Delta}{2},\l\rangle,\eea
so that the charge density operator appears.
Multiplying this expression by $e_q$,  summing over quark flavors $q$ and using  \eq{fdef}  and \eq{rhob0} shows that
\bea \sum_q e_q  \int dx\, d^2k\rho_q(\bfb,{\bf k},x,\l )=\r(b).\label{relate}\eea
This means that the transverse charge density exhibits a specific aspect of quark Wigner distributions. Observe that  proton electromagnetic 
    form factors    occupy  a small, but important, corner of a vast field. 
 \\

\section{True {\it vs.} Non-relativistic density}

\begin{figure}[h]
\centering
\includegraphics[scale = 0.5]{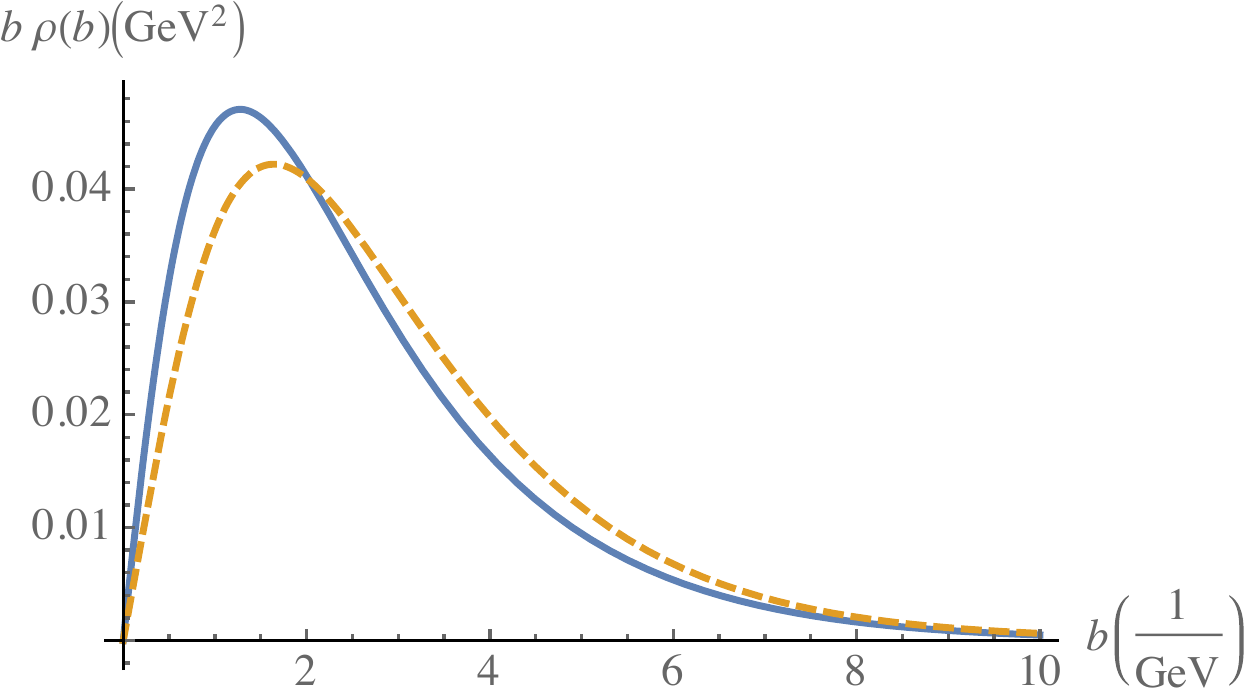}
\caption{True  (solid) vs non-relativistic (dashed) density}
\label{rhoPlot}\end{figure}

The  quantity $\r(b)$ of \eq{rhob0} is a true density. It is properly defined as the matrix element of a density operator between identical initial and final states.
It depends only on  the two-dimensional  transverse variable $\bfb$ because boosts in the longitudinal momentum depend on interactions. Furthermore, it  depends  only on the magnitude $b$  because  of its  independence of $\l$.\\

The purpose of this Section is to understand the differences between $\r(b)$ and $\rho_{\rm NR}(r)$ even though these are inherently different.  The former is well-defined in quantum field theory, and  the latter is defined as a three-dimensional Fourier transform of $G_E$.
However, it is useful to compare $\r(b)$ with a two-dimensional version of $\rho_{\rm NR}(r)$. A transverse non-relativistic density can be defined in analogy with the transverse densities that are  used in relativistic  heavy ion physics, and which generally appear in Glauber theory (eikonal approximation)  calculations of scattering processes. To this end, one writes $\vec r=  z\,\widehat{\bf k} +\bfb$ and then integrates  $\rho_{\rm NR}(r)$ over all values of $z$.
Then one obtains a  non-relativistic transverse density
\bea \rho_{\rm NR, T}(b)=\int_{-\infty}^\infty dz \, \rho_{\rm NR}(\sqrt{b^2+z^2})\label{zint}\eea
The use of \eq{rhoNR} in this equation followed by integration over angles yields the result
\bea  \rho_{\rm NR,T}(b)={1\over 2\pi}\int_0^\infty dQ Q J_0(b Q)G_E(Q^2),\eea
where $J_0$ is a cylindrical Bessel function. The non-relativistic transverse density  $ \rho_{\rm NR, T}(b)$ may be compared with
a more detailed version of the  true transverse density of \eq{rhob0}:
\bea  \rho(b)={1\over 2\pi}\int_0^\infty dQ Q J_0(b Q)F_1(Q^2).\eea 
The difference is simple: to obtain a relativistic transverse charge density one need only replace $G_E$ by $F_1$.
\\

Next specific values of $F_1$ and $G_E$ are used to obtain an explicit comparison.
The form factors $G_{E,M}$ have been measured \cite{Arrington:2006zm,Punjabi:2015bba}. Thus the quantity $F_1$ is readily available from the relation
\bea F_1(Q^2)={G_E(Q^2)+ {Q^2\over 4M^2} G_M(Q^2)\over 1+{Q^2\over 4M^2} }.\eea\\
Then the  functions $G_E$ and $F_1$,  obtained from  a recent parameterization \cite{Ye:2017gyb}, are   used to obtain 
the densities shown  in Fig.~\ref{rhoPlot}. The non-relativistic transverse density is seen to be very different  from the correct transverse density. In particular,  the non-relativistic version has a larger spatial extent.  \\

The spatial extent can be understood by computing the average value of $b^2$:
\bea
\la b^2\ra_{\rm NR}=\int d^2b \,b^2\r_{\rm NR}(b)=-4G_E'(0)={2\over3}r_p^2={2\over3}\la r^2\ra_{\rm NR}\\
\la b^2\ra=\int d^2b\, b^2\r(b)=-4F_1'(0)
\eea
The use of the relation \eq{GEdef} leads to the result:
\bea \la b^2\ra_{\rm NR}=\la b^2\ra +{\kappa\over 4M^2}= \la b^2\ra+0.02 \,{\rm fm}^2 \label{b2}
\eea
The difference between the true value and the non-relativistic one is very significant on the scale of distances relevant to the proton radius puzzle.  But to be clear: hydrogen spectroscopy  measures  the slope of $G_E$ at its 
origin and relates that quantity to $r_p^2$.
\subsection{Relativistic Moment Expansion- RME}

A moment expansion analogous to \eq{nrmom} can be derived from the relation between $F_1(Q^2)$ and the true density $\r(b)$.  Invert \eq{rhob0} to obtain
\bea F_1(Q^2)=\int d^2b \r(b) e^{i\bfQ\cdot\bfb}.
\eea
Then expand the exponential in a power series in $i\bfQ\cdot\bfb$ and also expand $F_1(Q^2)$ in powers of $Q^2$.
Equating the two expansions gives the result
\bea F_1(Q^2)=\sum_{n=0}^\infty (-1)^n{I_n\over (2n)!} \la b^{2n}\ra Q^{2n} F_1^{(n)}(0)\label{RME}\\
I_n\equiv {1\over2\pi} \int_0^{2\pi} d\phi \cos^{2n}\phi={(2n-1)!!\over 2^n \,n!}\\
\la b^{2n}\ra\equiv \int d^2b \r(b) b^{2n},
\eea
where the notation $F_1^{(n)}(0)$ denotes taking the $n$'th derivative of $F_1$ with respect to $Q^2$ at $Q^2=0$
The first terms are given by
\bea F_1(Q^2)\approx 1-{Q^2\over4}\la b^2\ra F^{(1)}(0) +{Q^4\over64}\la b^4\ra F_1^{(2)}(0)+\cdots \,.\eea
Each of the moments $\la b^{2n}\ra$  is invariant under Lorentz transformations.  \\

Hydrogen spectroscopy depends on the moments $\la r^{2,4}\ra_{\rm NR}$.
The RME of \eq{RME} can  be used to determine  the true moments $\la b^{2n}\ra $ in terms of the non-relativistic ones.   The use of \eq{b2} leads to
the result:
\bea \la b^2\ra={2\over3}\la r^2\ra_{\rm NR}-{\kappa\over 4M^2},\eea
and 
\bea  \la b^4\ra={8\over15}\la r^4\ra_{\rm NR}+{8\over 3M^2} \la r^2\ra_{\rm NR}  -{4\m\over 3M^2} \la r^2_M\ra_{\rm NR}-{4\kappa\over M^4},\eea
 where $\m=1+\kappa,\, \la r^2_M\ra_{\rm NR}\equiv -6G_M'(0)$.\\
 
 The moments of $b$ are closely tied to GPDs, which are accessible experimentally and through lattice calculations.
 \section{Summary} 
 
 This paper 
 unites  the hydrogen spectroscopy literature   with that of lepton-proton scattering  to  show how \eq{basic} emerges from the separate literatures.\\
 
 The appearance  of the proton radius $r_p$  in hydrogen spectroscopy  is discussed in Sect.II, which shows that the energy shift caused by the non-zero extent of the proton (\eq{good})  is proportional to the slope of $G_E(Q^2)$ at its origin. An explicit three-dimensional  charge density does not appear.  There is no need to define $r_p^2$ as a moment of such a density. \\
 
 Sect.~III begins with a brief historical review of how a non-relativistic, frame-dependent  spherically-symmetric, three-dimensional charge density    was postulated in the early work of Hofstadter and co-workers.  There is only one attempted derivation of this density in the literature \cite{Sachs:1962zzc}. This derivation is shown to be faulty because it used states of completely uncertain position, which leads to an infinite contribution, \eq{M21}. \\
 
 A properly defined relativistic three-dimensional charge density is discussed in Sect.~IV. This quantity is intimately connected with  modern formulations of the diverse set of possible parton distributions. It depends on longitudinal and   transverse momentum or longitudinal and transverse   position. The two-dimensional transverse density $(\r(b))$, obtained as an integral over the longitudinal coordinate, is a two-dimensional Fourier transform of the Dirac form factor $F_1$, \eq{rhob0}.  The transverse density is shown to be related to  specific integrals of a Wigner distribution, \eq{relate}.\\
 
 The phenomenology of $\r(b)$ is discussed in Sect.~V.  The non-relativistic version is shown to be significantly different from the correct density, and a  correctly defined moment expansion $\rm RME$ is derived, \eq{RME}.  These moments are related to the  non-relativistic ones. \\
 
  \section*{Appendix A-Derivation of \eq{key}}
 The expression $\bar u(p',s')\G^0u(p,s)$  (\eq{vert} with $\vec p'=\vec p+\vq$) is evaluated here. The spinor is given  to order $1/M^2$ by
 \bea u^\dagger(p,s)=[1\quad {\vec{\s}\cdot \vec p\over 2m}]/(1+\vp^2/(8M^2)),
 \eea  
 in which the spin-dependent term is neglected.
 Then first evaluate \bea
 & \bar u(p',s')\g^0u(p,s)F_1=(1+ {\vec{\s}\cdot \vec p^{\,'}\over 2M}{\vec{\s}\cdot \vec p\over 2M})(1-{\vp'^2+\vp^2\over 8M^2})F_1\\&=(1-{\vq^2\over 8M^2})F_1.\label{g0}
  \eea
 Next evaluate the term proportional to $F_2$. Use $i\s^{0\n}q_n=ii{1\over 2}[\g^0,-\vec{\s}\cdot\vq]=\g^0\vec{\g}\cdot\vq$,
 so that
 \bea  \bar u(p',s') {i\s^{0\n}q_\n \over 2M}u(p,s)F_2= [\vec\s\cdot\vq\vec\s\cdot\vp-\vec\s\cdot(\vp+\vq)\vec \s\cdot\vec q]F_2/(4M^2)=-{\vq^2\over 4M^2}F_2,\label{g2}\eea
 in which a spin-dependent term is omitted.
 Combining the results \eq{g0} and \eq{g2} and recalling the definition \eq{vert} leads immediately to \eq{key}.
 \section*{Appendix B-General Wave Packet }
Sect.~III showed that  the derivation of the relation between $G_E$ and a three-dimensional charge density was incorrect because it ignores an infinite term . A specific representation, \eq{gdef}, of the delta function was used to demonstrate this flaw.
 Here we show that an infinite term appears for any representation of the delta function. 
 The evaluation of \eq{M20} involves the combination
 $ \nabla_q^2 \int d^3p \,g(\vp+\vq/2)g^*(\vp-\vq/2)$.
 Other terms in which $\vec{\nabla}_q$ acts on  the product of the $g_R$ factors times the matrix element of $\G^0$ or only on $\G^0$ are not infinite. But this term is actually infinite.
 To see this introduce the Fourier transform:
 \bea g(\vp)=\int d^3x \tilde{g}(\vec {x}) e^{i\vp\cdot \vec {x}}.\eea
The function $g(\vp) $ is very narrow in momentum space, so $\tilde g(\vec x)$ must be very broad in coordinate space. Using  Fourier transforms one finds 
\bea  \nabla_q^2 \int d^3p\,g(\vp+\vq/2)g^*(\vp-\vq/2)= -(2\pi)^3\int d^3r\, r^2 |\tilde{g}(\vec r)|^2.\eea
 Thus it is immediately plausible that this  quantity is infinite for any representation of the delta function.\\
 
The infinite result can proved.  Let the general form, dictated by dimensional analysis,  of $g(\vp)$ be given by
\bea g_R(\vp)=R^{3/2} F(p R),\eea 
where $F$ is a dimensionless, real-valued function and we understand that the limit $R\to \infty$ is to  be taken after doing the relevant integrals.  
Then 
\bea \tilde g_R(\vec r)=\int {d^3p\over (2\pi)^3} e^{i\vp\cdot \vec r} g_R(\vp)={1\over 2\pi^2 R^{3/2}} \,{R\over r}\int_0^\infty du \,u \,\sin (u r/R)\, F(u)\equiv {1\over R^{3/2}} G(r/R),
\eea
and 
\bea \int d^3r \,r^2 | \tilde g_R(\vec r)|^2= {1\over R^3}\int d^3r r^2 |G(r/R)|^2=R^2 \int d^3z z^2 |G(z)|^2,\eea
where $z$ is a dimensionless variable. The function $G$ is normalizable, and its $z^2$-weighted integral is finite, because it is a Fourier transform of a normalizable function. However, the entire expression is proportional to $R^2$ which is infinite.\\

The net result is that the wave packet treatment of Sachs is not feasible, no matter how a delta function is represented.\\

\section*{Acknowledgments}
This work was stimulated by a suggestion from J. Bernauer. I thank E.~Voutier for a useful discussion.
This work was supported by the U. S. Department of Energy Office of Science, Office of Nuclear Physics under Award Number DE-FG02-97ER-41014. 


\begin{thebibliography}{99}

\bibitem{pohl:2010zza} 
  R.~Pohl, A.~Antognini, F.~Nez, F.~D.~Amaro, F.~Biraben, J.~M.~R.~Cardoso, D.~S.~Covita and A.~Dax {\it et al.},
  Nature {\bf 466}, 213 (2010).

 \bibitem{Antognini:1900ns} 
  A.~Antognini, F.~Nez, K.~Schuhmann, F.~D.~Amaro, FrancoisBiraben, J.~M.~R.~Cardoso, D.~S.~Covita and A.~Dax {\it et al.},
  Science {\bf 339}, 417 (2013).
  
  \bibitem{Horbatsch:2015qda} 
  M.~Horbatsch and E.~A.~Hessels,
  Phys.\ Rev.\ C {\bf 93}, 015204 (2016)
  
  \bibitem{Miller:2011yw} 
  G.~A.~Miller, A.~W.~Thomas, J.~D.~Carroll and J.~Rafelski,
  Phys.\ Rev.\ A {\bf 84}, 020101 (2011)

\bibitem{Pohl:2013yb} 
  R.~Pohl, R.~Gilman, G.~A.~Miller and K.~Pachucki,
  Ann.\ Rev.\ Nucl.\ Part.\ Sci.\  {\bf 63}, 175 (2013)
  
\bibitem{Carlson:2015jba} 
  C.~E.~Carlson,
  Prog.\ Part.\ Nucl.\ Phys.\  {\bf 82}, 59 (2015)
  
  \bibitem{beyer:2017}
A.~Beyer {\it et al.}, Science, 358, 79 (2017)
%

\bibitem{Fleurbaey:2018fih} 
  H.~Fleurbaey {\it et al.},
  Phys.\ Rev.\ Lett.\  {\bf 120}, no. 18, 183001 (2018)

\bibitem{Gasparian:2014rna} 
  A.~Gasparian [PRad at JLab Collaboration],
  EPJ Web Conf.\  {\bf 73}, 07006 (2014). 

\bibitem{Gilman:2013eiv} 
  R.~Gilman {\it et al.} [MUSE Collaboration],
  arXiv:1303.2160 [nucl-ex].



\bibitem{BMot}
A. Bohr and B. R.  Mottelson, ``Nuclear Structure,, Volume I
W. A. Benjamin, Inc. New York, Amsterdam 1969

\bibitem{PB}
M.~A.~Preston and R.~K.~Bhaduri, ``Structure of the nucleus",
Addison-Wesley, Reading USA 1975

\bibitem{Frauenfelder:1979ii} 
  H.~Frauenfelder and E.~M.~Henley,
  ``Subatomic Physics,''
  Prentice-Hall, Englewood, USA 1974,1991
  
\bibitem{Cheng:1979ay} 
  D.~C.~Cheng and G.~K.~O'Neill,
  Reading, Usa: Addison-wesley (1979) 423p
  
\bibitem{Halzen:1984mc} 
  F.~Halzen and A.~D.~Martin,
  ``Quarks And Leptons: An Introductory Course In Modern Particle Physics,''
  New York, Wiley ( 1984) 
  

\bibitem{Wong:1998ex} 
  S.~S.~M.~Wong,
  New York, USA: Wiley (1998) 
  
\bibitem{Thomas:2001kw} 
  A.~W.~Thomas and W.~Weise,
  ``The Structure of the Nucleon,''
  Berlin, Germany: Wiley-VCH (2001) 389 p
  
\bibitem{Close:2007zzd} 
 R. Ent, page 39 of  F.~Close, S.~Donnachie and G.~Shaw,
  Camb.\ Monogr.\ Part.\ Phys.\ Nucl.\ Phys.\ Cosmol.\  {\bf 25}, 1 (2007).
  
\bibitem{Hofstadter:1956qs} 
  R.~Hofstadter,
  Rev.\ Mod.\ Phys.\  {\bf 28}, 214 (1956).
  \bibitem{Hofstadter:1958} R.~Hofstadter, F.~Bumiller, and M.~R.~Yearian, Rev. Mod. Phys. {\bf 30}, 482 (1958).
\bibitem{RHN} R.~Hofstadter, "The Electron Scattering Method \& its Application to the Structure of Nuclei and Nucleons," Nobel Lectures, Physics 1942Ð1962, pp. 560Ð581, Elsevier Pub. Co., Amsterdam-London-New York (Dec 1961).


\bibitem{Boer:2011fh} 
  D.~Boer {\it et al.},
  arXiv:1108.1713 [nucl-th].
  
  \bibitem{Dudek:2012vr} 
  J.~Dudek {\it et al.},
  Eur.\ Phys.\ J.\ A {\bf 48}, 187 (2012)
  
  \bibitem{Ji:2003ak} 
  X.~D.~Ji,
  Phys.\ Rev.\ Lett.\  {\bf 91}, 062001 (2003)

\bibitem{Wigner:1932eb} 
  E.~P.~Wigner,
  Phys.\ Rev.\  {\bf 40}, 749 (1932).

\bibitem{Mueller:1998fv} 
  D.~Mueller, D.~Robaschik, B.~Geyer, F.-M.~Dittes and J.~Horejsi,
  Fortsch.\ Phys.\  {\bf 42}, 101 (1994)
 
  \bibitem{Ji:1996nm} 
  X.~D.~Ji,
  Phys.\ Rev.\ D {\bf 55}, 7114 (1997)
\bibitem{Ji:1996ek} 
  X.~D.~Ji,
  Phys.\ Rev.\ Lett.\  {\bf 78}, 610 (1997)

\bibitem{Radyushkin:1996nd} 
  A.~V.~Radyushkin,
  Phys.\ Lett.\ B {\bf 380}, 417 (1996)
\bibitem{Radyushkin:1997ki} 
  A.~V.~Radyushkin,
  Phys.\ Rev.\ D {\bf 56}, 5524 (1997) 
\bibitem{Collins:1996fb} 
  J.~C.~Collins, L.~Frankfurt and M.~Strikman,
  Phys.\ Rev.\ D {\bf 56}, 2982 (1997) 
 
\bibitem{Ji:1998pc}
X.-D. Ji,
 J. Phys. {\bf G24}, 1181 (1998), hep-ph/9807358.

\bibitem{Radyushkin:2000uy}
A.~V. Radyushkin, ``Generalized parton distributions,''
  In Shifman, M. (ed.): At the frontier of particle physics, vol. 2, 1037-1099
 (2000), hep-ph/0101225.

\bibitem{Goeke:2001tz}
K.~Goeke, M.~V. Polyakov, and M.~Vanderhaeghen,
\newblock Prog. Part. Nucl. Phys. {\bf 47}, 401 (2001), hep-ph/0106012.

\bibitem{Diehl:2003ny}
M.~Diehl,
\newblock Phys. Rept. {\bf 388}, 41 (2003), hep-ph/0307382.

\bibitem{Ji:2004gf}
X.~Ji,
 Ann. Rev. Nucl. Part. Sci. {\bf 54}, 413 (2004).

\bibitem{Belitsky:2005qn}
A.~V. Belitsky and A.~V. Radyushkin,
\newblock Phys. Rept. {\bf 418}, 1 (2005), hep-ph/0504030.

\bibitem{Burkardt:2002hr} 
  M.~Burkardt,
  Int.\ J.\ Mod.\ Phys.\ A {\bf 18}, 173 (2003)
\bibitem{Diehl:2002he} 
  M.~Diehl,
  Eur.\ Phys.\ J.\ C {\bf 25}, 223 (2002)
  Erratum: [Eur.\ Phys.\ J.\ C {\bf 31}, 277 (2003)]

\bibitem{Ralston:1979ys}
J.~P. Ralston and D.~E. Soper,
\newblock Nucl. Phys. {\bf B152}, 109 (1979).
\bibitem{Mulders:1995dh}
P.~J. Mulders and R.~D. Tangerman,
\newblock Nucl. Phys. {\bf B461}, 197 (1996), hep-ph/9510301.

 \bibitem{Belitsky:2002sm}
A.~V. Belitsky, X.~Ji, and F.~Yuan,
Nucl. Phys. {\bf B656}, 165 (2003), hep-ph/0208038.

  \bibitem{Miller:2008sq} 
G.~A.~Miller,
  Phys.\ Rev.\ C {\bf 68}, 022201 (2003);
  G.~A.~Miller,
  Nucl.\ Phys.\ News {\bf 18}, 12 (2008); G.~A.~Miller,
  Phys.\ Rev.\ C {\bf 76}, 065209 (2007)


\bibitem{Collins:1981uk} 
  J.~C.~Collins and D.~E.~Soper,
  Nucl.\ Phys.\ B {\bf 193}, 381 (1981)
  Erratum: [Nucl.\ Phys.\ B {\bf 213}, 545 (1983)].

 
  
  \bibitem{Boer:1997nt} 
  D.~Boer and P.~J.~Mulders,
  Phys.\ Rev.\ D {\bf 57}, 5780 (1998)

 \bibitem{Eides:2000xc} 
  M.~I.~Eides, H.~Grotch and V.~A.~Shelyuto,
  Phys.\ Rept.\  {\bf 342}, 63 (2001)
  doi:10.1016/S0370-1573(00)00077-6

\bibitem{Eides:2007xc} 
  M.~I.~Eides, H.~Grotch and V.~A.~Shelyuto,
  ``Theory of light hydrogenic  - like atoms,''
 Springer-Verlag Berlin Heidelberg 2007

\bibitem{Eides:2014swa} 
  M.~I.~Eides,
  ``Recent ideas on the calculation of lepton anomalous magnetic moments,''
  [arXiv:1402.5860 [hep-ph]].

\bibitem{Breit:1929zz} 
  G.~Breit,
  Phys.\ Rev.\  {\bf 34}, 553 (1929).
\bibitem{Breit:1930zza} 
  G.~Breit,
  Phys.\ Rev.\  {\bf 36}, 383 (1930).

\bibitem{Barker:1955zz} 
  W.~A.~Barker and F.~N.~Glover,
  Phys.\ Rev.\  {\bf 99}, 317 (1955).

\bibitem{Bertozzi:1972jff} 
  W.~Bertozzi, J.~Friar, J.~Heisenberg and J.~W.~Negele,
  Phys.\ Lett.\  {\bf 41B}, 408 (1972).
  
  \bibitem{Stryker:2015ika} 
  J.~R.~Stryker and G.~A.~Miller,
  Phys.\ Rev.\ A {\bf 93}, no. 1, 012509 (2016)

  \bibitem{Preedom:1987mx} 
  B.~M.~Preedom and R.~Tegen,
  Phys.\ Rev.\ C {\bf 36}, 2466 (1987).

\bibitem{Rosenbluth:1950yq} 
  M.~N.~Rosenbluth,
  Phys.\ Rev.\  {\bf 79}, 615 (1950).
  
  
   \bibitem{Ernst:1960zza} 
  F.~J.~Ernst, R.~G.~Sachs and K.~C.~Wali,
  Phys.\ Rev.\  {\bf 119}, 1105 (1960).



\bibitem{Sachs:1962zzc} 
  R.~G.~Sachs,
  Phys.\ Rev.\  {\bf 126}, 2256 (1962).

\bibitem{Miller:2009qu} 
  G.~A.~Miller,
  Phys.\ Rev.\ C {\bf 79}, 055204 (2009)
 
 
  \bibitem{Gasiorowicz:1966xra} 
  S.~Gasiorowicz,
  ``Elementary particle physics,''
John Wiley \& Sons, New York 1966

\bibitem{Cahill:2013pez} 
  K.~Cahill,
  ``Physical Mathematics,''
  Cambridge University Press, New York 2013

\bibitem{Born:1999ory} 
  M.~Born and E.~Wolf,
  ``Principles of Optics: Electromagnetic Theory of Propagation, Interference and Propagation of Light,''
  7th Edition Appendix IV
  Cambridge University Press Cambridge, 1999
   \bibitem {AW2005} G.~B.Arfken and H.~J.~Weber,
 ``Mathematical Methods For Physicists",
 Elsevier, Amsterdam 2005
 \bibitem{Susskind:1968zz} 
  L.~Susskind,
  Phys.\ Rev.\  {\bf 165}, 1547 (1968).

\bibitem{Miller:2010nz} 
  G.~A.~Miller,
  Ann.\ Rev.\ Nucl.\ Part.\ Sci.\  {\bf 60}, 1 (2010) \\ 
  G.~A.~Miller,
  Phys.\ Rev.\ Lett.\  {\bf 99}, 112001 (2007)


  \bibitem{Brodsky:2000ii} 
  S.~J.~Brodsky, D.~S.~Hwang, B.~Q.~Ma and I.~Schmidt,
  Nucl.\ Phys.\ B {\bf 593}, 311 (2001)
  
\bibitem{Miller:2009sg} 
  G.~A.~Miller,
  Phys.\ Rev.\ C {\bf 80}, 045210 (2009)
     \bibitem{Soper:1976jc} 
  D.~E.~Soper,
  Phys.\ Rev.\ D {\bf 15}, 1141 (1977).
  

  \bibitem{Carlson:2007xd} 
  C.~E.~Carlson and M.~Vanderhaeghen,
  Phys.\ Rev.\ Lett.\  {\bf 100}, 032004 (2008)
\bibitem{McLerran:1993ka} 
  L.~D.~McLerran and R.~Venugopalan,
  Phys.\ Rev.\ D {\bf 49}, 3352 (1994)
  \bibitem{Dumitru:2018vpr} 
  A.~Dumitru, G.~A.~Miller and R.~Venugopalan,
  Phys.\ Rev.\ D {\bf 98}, no. 9, 094004 (2018)
  \bibitem{Hagler:2004yt}
P.~Hagler,
\newblock Phys. Lett. {\bf B594}, 164 (2004), hep-ph/0404138.

\bibitem{Boffi:2007yc}
S.~Boffi and B.~Pasquini,
\newblock Riv. Nuovo Cim. {\bf 30}, 387 (2007), 0711.2625.

\bibitem{Collins:1981uw}
J.~C. Collins and D.~E. Soper,
\newblock Nucl. Phys. {\bf B194}, 445 (1982).

\bibitem{Belyaev:1988xu}
V.~M. Belyaev and B.~L. Ioffe,
\newblock Nucl. Phys. {\bf B313}, 647 (1989).

\bibitem{Anselmino:1994gn}
M.~Anselmino, A.~Efremov, and E.~Leader,
\newblock Phys. Rept. {\bf 261}, 1 (1995), hep-ph/9501369.


\bibitem{Chibisov:1995ss}
B.~Chibisov and A.~R. Zhitnitsky,
\newblock Phys. Rev. {\bf D52}, 5273 (1995), hep-ph/9503476.

\bibitem{Pasquini:2009eb}
B.~Pasquini, S.~Boffi, A.~V. Efremov, and P.~Schweitzer,
\newblock (2009), 0912.1761.
 
 
 \bibitem{Echevarria:2016mrc} 
  M.~G.~Echevarria, A.~Idilbi, K.~Kanazawa, C.~Lorcé, A.~Metz, B.~Pasquini and M.~Schlegel,
  Phys.\ Lett.\ B {\bf 759}, 336 (2016)
  

  
  
  \bibitem{Lorce:2011kd} 
  C.~Lorce and B.~Pasquini,
  Phys.\ Rev.\ D {\bf 84}, 014015 (2011)
     
       \bibitem{Accardi:2012qut} 
  A.~Accardi {\it et al.},
  Eur.\ Phys.\ J.\ A {\bf 52}, 268 (2016)
     \bibitem{Kogut:1969xa} 
  J.~B.~Kogut and D.~E.~Soper,
  Phys.\ Rev.\ D {\bf 1}, 2901 (1970).
  
  \bibitem{Burkardt:2005td} 
  M.~Burkardt,
  Int.\ J.\ Mod.\ Phys.\ A {\bf 21}, 926 (2006)
  
  \bibitem{Burkardt:2000za} 
  M.~Burkardt,
  Phys.\ Rev.\ D {\bf 62}, 071503 (2000)
  \bibitem{Diehl:2000xz} 
  M.~Diehl, T.~Feldmann, R.~Jakob and P.~Kroll,
  Nucl.\ Phys.\ B {\bf 596}, 33 (2001)
  Erratum: [Nucl.\ Phys.\ B {\bf 605}, 647 (2001)]

\bibitem{Belitsky:2003nz} 
  A.~V.~Belitsky, X.~d.~Ji and F.~Yuan,
  Phys.\ Rev.\ D {\bf 69}, 074014 (2004)
 
\bibitem{Meissner:2009ww} 
  S.~Meissner, A.~Metz and M.~Schlegel,
  JHEP {\bf 0908}, 056 (2009)


 

   \bibitem{Arrington:2006zm} 
  J.~Arrington, C.~D.~Roberts and J.~M.~Zanotti,
  J.\ Phys.\ G {\bf 34}, S23 (2007)
\bibitem{Punjabi:2015bba} 
  V.~Punjabi, C.~F.~Perdrisat, M.~K.~Jones, E.~J.~Brash and C.~E.~Carlson,
  Eur.\ Phys.\ J.\ A {\bf 51}, 79 (2015)
\bibitem{Ye:2017gyb} 
  Z.~Ye, J.~Arrington, R.~J.~Hill and G.~Lee,
  Phys.\ Lett.\ B {\bf 777}, 8 (2018)
  



 






  



    
  
  
    


 



  \end{thebibliography}

 \end{document}